\pdfoutput=1
\documentclass[twocolumn,aps,prb,superscriptaddress]{revtex4-1}
  \usepackage{graphicx,wrapfig,amsmath,upgreek,mathtools,amsfonts,physics,multirow,color}
  \usepackage{graphicx}
  \usepackage{color}
  \usepackage{bbold}
  \usepackage{amsmath}
  \usepackage{enumitem}
  \usepackage{amssymb}
  \usepackage{hyperref}

  \newcommand{\be}{\begin{equation}}
  \newcommand{\ee}{\end{equation}}
  
  \newcommand{\bea}{\begin{eqnarray}}
  \newcommand{\eea}{\end{eqnarray}}

  \newcommand{\la}{\left\langle}
  \newcommand{\ra}{\right\rangle}

  \newcommand{\lp}{\left(}
  \newcommand{\rp}{\right)}

  \renewcommand{\vec}[1]{{\boldsymbol #1}}
  \def\nn{\nonumber\\}

  \newenvironment{nalign}{
    \begin{equation}
    \begin{aligned}
}{
    \end{aligned}
    \end{equation}
}

\begin{document}
  \title{Dipole-active collective excitations in moir\'e flat bands}
  \author{Ali Fahimniya} 
  \thanks{These authors contributed equally to this work.}
  \affiliation{Massachusetts Institute of Technology, Cambridge, Massachusetts 02139, USA}
  \author{Cyprian Lewandowski}
  \thanks{These authors contributed equally to this work.}
    \affiliation{Massachusetts Institute of Technology, Cambridge, Massachusetts 02139, USA}
\affiliation{Department of Physics, California Institute of Technology, Pasadena, CA 91125, USA}
  \author{Leonid Levitov}
  \affiliation{Massachusetts Institute of Technology, Cambridge, Massachusetts 02139, USA}

  \begin{abstract}
Collective plasma excitations in moir\'e flat bands display unique properties reflecting strong electron-electron interactions and unusual carrier dynamics in these systems. Unlike the conventional two-dimensional plasmon modes, dispersing as $\sqrt{k}$ at low frequencies and plunging into particle-hole continuum at higher frequencies, the moir\'e plasmons pierce through the flat-band continuum and acquire a strong over-the-band character. Due to the complex structure of the moir\'e superlattice unit cell, the over-the-band plasmons feature several distinct branches connected through zone folding in the superlattice Brillouin zone. Using a toy Hubbard model for the correlated insulating order in a flat band, we predict that these high-frequency modes become strongly dipole-active upon the system undergoing charge ordering, with the low-frequency modes gapped out within the correlated insulator gap. Strong dipole moments and sensitivity to charge order make these modes readily accessible by optical measurements, providing a convenient diagnostic of the correlated states.
  
  \end{abstract}
  \date{\today}
  \maketitle

\section{Introduction}

The moir{\'e} materials, such as twisted bilayer graphene (TBG), host narrow electron bands populated by strongly interacting carriers which exhibit a variety of ordered correlated states\cite{cao2018_1,cao2018_2}. 
The unique properties of these materials call for reexamining some of the long-standing questions in many-body physics of low-dimensional systems. One intriguing question concerns the properties of the correlated insulating (CI) states\cite{efetov2019,serlin2019intrinsic,Sharpe605,SNP19, Zondiner2020, Wong2020, Balents2020}, in particular the dynamical response and the properties of collective modes in these states.

Collective modes in strongly interacting electron systems were thoroughly studied in the past, theoretically and experimentally, in the context of ordered states in two-dimensional (2D) systems such as electrons on He surface and in semiconductor heterojunctions. 
In the regime of low carrier densities, such that Coulomb interactions are much stronger than the kinetic energy of the electron zero-point motion, 
these systems exhibit long-range ordered states described as a Wigner solid. The collective modes in a Wigner solid can be understood as longitudinal and transverse phonons with the spectrum derived from the long-range $1/r$ potential\cite{PhysRevA.8.2136,PhysRevB.10.3150,PhysRevB.15.1959}. The long-wavelength dispersion of longitudinal modes 
of the electron crystal is identical to that of plasmons in a two-dimensional electron gas, $\omega=\beta \sqrt{k}$, $\beta^2=2\pi e^2 n / m$ where $e$, $m$ and $n$ are the carrier charge, mass and density; as a result, the longitudinal mode has little knowledge about charge order. The transverse mode, to the contrary, exists only in the crystal state but not in a gas; it has a linear dispersion $\omega=\beta' k$ that depends on the specifics of the crystalline order \cite{PhysRevA.8.2136,PhysRevB.10.3150,PhysRevB.15.1959,stern1967,1972JETP...35..395C,PhysRevB.9.4724}.

\begin{figure}[!t]
\centering
\includegraphics[width=\linewidth]{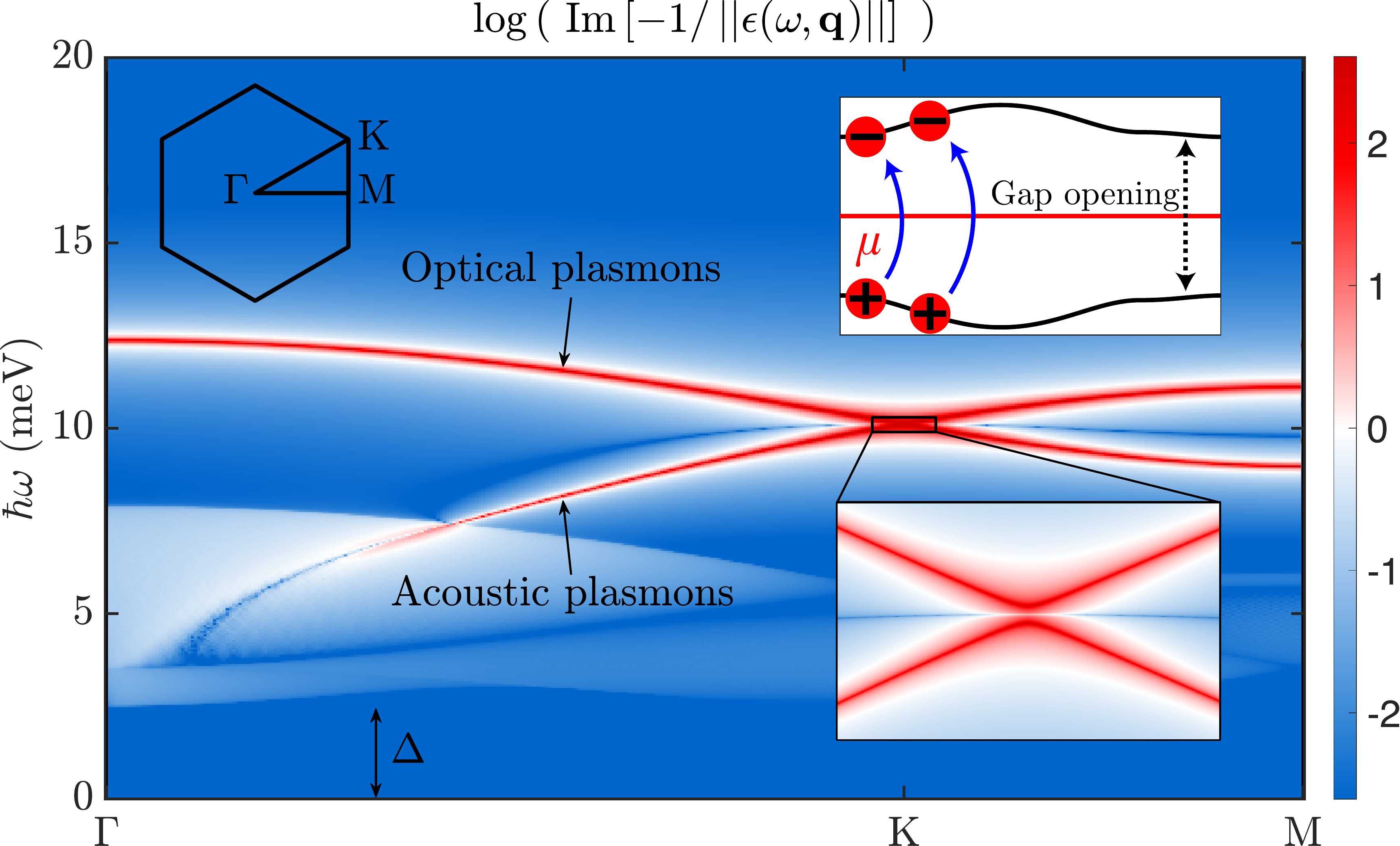}
\caption{Electron loss function for a narrow-band toy model of a correlated insulator, a tight-binding band in hexagonal lattice with carrier density at charge neutrality and spontaneously broken A/B sublattice symmetry. Broken symmetry gives rise to a correlation-driven gap $\Delta$ opening, Eq.~\eqref{eq:main_uu'}, in the electron bandstructure (see upper-right inset). The acoustic plasmon mode becomes gapped at energies below $\Delta$. The diatomic structure (two sublattices) gives rise to a second plasmon mode with optical-like dispersion at small momenta. As detailed in Fig.~\ref{fig:fig_4}, this mode features a non-zero dipole moment which is enhanced in the symmetry-broken CI state. The A/B sublattice symmetry breaking also gives rise to an avoided band crossing of the acoustic and optical modes at the zone boundary (lower inset). Parameter values are chosen to mimic TBG bands (bandwidth $W = 3.75$~meV, lattice periodicity $L_M=13.4$~nm). The CI energy gap is set at $\Delta=2W/3$; log scale is used to clarify the relation between different features. }
\label{fig:fig_1}
\end{figure}

Collective plasma excitations in narrow-band materials, such as TBG, have interesting properties that set them apart from the conventional 2D plasmons\cite{Lewandowski20869,PhysRevB.102.125408,PhysRevB.102.125403,doi:10.1021/acs.nanolett.6b02587,hu2017,2019arXiv191007893H}. One new aspect, first pointed out in Ref.\onlinecite{Lewandowski20869}, is that plasmon modes have strong dispersion that allows them to pierce through the particle-hole continuum to emerge above it, becoming decoupled from particle-hole excitations. This behavior is much unlike that seen in other two-dimensional systems, where plasmon modes plunge into the particle-hole continuum, merging with it and becoming damped\cite{stern1967,wunsch,hwang,jablan}. 
In contrast, the low-frequency behavior of these modes resembles that of the conventional 2D plasmons: a square-root dispersion $\omega=\beta \sqrt{k}$ with the stiffness $\beta$ that depends, as a square root, on the energy scale for the electronic transitions responsible for the collective mode - here set by the width of the narrow band\cite{Lewandowski20869,PhysRevB.102.125408,PhysRevB.102.125403}. 

Here we analyze collective modes in a correlated insulator state using a Hubbard toy model framework\cite{mahan2000many-particle} which mimics the essential aspects of the more microscopic models\cite{Vishwanath2019_CI_1,Vafek_2019_CI,Zaletel_2020_CI_1,MacDonald_2020_CI_1,Guinea_2020_CI_1}. We find that the ``over-the-band'' character of the plasmon modes remains largely unaltered by the charge-ordered state. This is illustrated in Fig.~\ref{fig:fig_1} which shows plasmon modes (bright red features) for a CI state in a tight-binding band in a honeycomb lattice with carrier density at charge neutrality (see schematic in the inset). The dispersion at small momenta depends on the nature of the correlated state, becoming gapped when the system is insulating. At large momenta, however, the behavior is essentially $\omega=\beta\sqrt{k}$ with only subtle details being sensitive to the low-energy electronic excitations in the narrow band.

The origin of the over-the-band can be elucidated by considering an analogy with the collective modes in a Wigner solid. The latter feature longitudinal modes dominated by the long-range $1/r$ interactions that are pushed far above the transverse modes dominated by the short-range vibrational degrees of freedom\cite{PhysRevA.8.2136,PhysRevB.10.3150,PhysRevB.15.1959,stern1967,1972JETP...35..395C,PhysRevB.9.4724}. A similar picture applies to the moir\'e flat bands, where the short-range and long-range lengthscales are defined relative to the moir\'e superlattice periodicty. The short-range degrees of freedom in this system are represented by the continuum of particle-hole excitations; their energies are essentially unaffected by the long-range $1/r$ interactions and therefore lie well below the longitudinal collective modes, labeled ``acoustic plasmons" in Fig.\ref{fig:fig_1}.

In addition to these modes, a new optical plasmon mode appears above the longitudinal branch (see Fig.~\ref{fig:fig_1}). Its origin can be understood, in a simplest way, by noting that the TBG superlattice is not ``monoatomic''. The matrix character of the TBG Hamiltonian makes the problem equivalent to the tight-binding problem on a diatomic lattice which exhibits a mode doubling behavior. Therefore, the optical plasmon mode is expected to be present in TBG bands irrespective of the occurrence of the A/B symmetry breaking and CI states. However, the change in spatial periodicity in the presence of spatial modulation accompanying CI orders can give rise, through zone folding, to additional optical plasmon modes. Even more importantly, where the presence of the CI state will matter most is in the dipole moment of the optical modes - as we will see the dipole magnitude grows sharply upon the system transitioning into the symmetry broken state. 

As a side remark, recent work \cite{doi:10.1021/acs.nanolett.6b02587} analyzed
and measured\cite{2019arXiv191007893H} plasmon excitations originating from transitions between flat and dispersive bands in TBG. These transitions, as well as the resulting modes, are distinct from the intra-flat-band transitions analyzed here, which are 
depicted by blue arrows in the upper inset of Fig.~\ref{fig:fig_1}. As discussed below, the intraband character of the resulting modes makes them particularly sensitive to charge ordering and correlated states in moir\'e bands.

\section{Correlated Hubbard insulator}

We begin by introducing a model of a correlated Hubbard insulator used for the analysis. We start with a narrow tight-binding band with an on-site Hubbard interaction
 \begin{align}\label{eq:H_main}
 H=\sum_{\la ii'\ra}\sum_\sigma t a^\dagger_{i\sigma}a_{i'\sigma}+\sum_i\sum_{\sigma\ne\sigma'} U n_{i\sigma}n_{i\sigma'}
 ,
 \end{align}
 with the interaction strength $U$ and hopping $t$ chosen to mimick the behavior of electrons in TBG narrow bands as discussed below. In general,  the spin variables $\sigma$ should account for both spin and valley degrees of freedom. Here, for illustration purposes, we suppress the valley degrees of freedom, taking spin variables to be $\sigma=\uparrow,\downarrow$. In which case the interaction term takes the usual Hubbard form $\sum_i Un_{i\uparrow}n_{i\downarrow}$. This suppression of valley degree of freedom will indirectly lead to a gap opening at a charge neutrality point of the electronic system instead of the expected TBG fillings\cite{cao2018_1,efetov2019}. Here we use this model as a simple framework to describe order in a correlated Hubbard insulator. For simplicity, we will treat hopping as nearest neighbor and take the lattice to be bipartite, with all lattice sites, in the absence of the order, having equal on-site energies. In this case, the ordered state can be described by site occupancies that differ on the even and odd sublattices:
\be
\bar n_{i\uparrow}=\begin{cases}\bar n, & {\rm even\ sites} \\ \bar n', & {\rm odd\ sites} \end{cases}
,\, 
\bar n_{i\downarrow}=\begin{cases}\bar n', & {\rm even\ sites} \\ \bar n, & {\rm odd\ sites} \end{cases}
\ee
The difference in the energies between the even and odd sites for one spin polarization represents an order parameter, arising spontaneously in the ordered state:
\be\label{eq:main_uu'}
\Delta =u-u'
,\,
u=Un_{i\uparrow}=Un_{i'\downarrow}
,\,
u'=Un_{i'\uparrow}=Un_{i\downarrow}
\ee
(here $i$ and $i'$ label the even and odd sublattice sites). The quantity $\Delta$ also represents the 
energy gap in the ordered state. 

To describe the ordered state and to estimate $\Delta$, we develop a mean field approximation. This is done in a standard way by writing 
\be
n_{i\sigma}=\bar n_{i\sigma}+\delta n_{i\sigma}
.
\ee
Plugging this expression in the Hubbard interaction we  expand it in fluctuations about the mean-field state:
 \begin{align}
& \sum_i Un_{i\uparrow}n_{i\downarrow}=\sum_i U(\bar n_{i\uparrow}+\delta n_{i\uparrow})(\bar n_{i\downarrow}+\delta n_{i\downarrow})
\nn
&=\sum_i U\bar n_{i\uparrow}\bar n_{i\downarrow}+U\delta n_{i\uparrow}\bar n_{i\downarrow}+U\bar n_{i\uparrow}\delta n_{i\downarrow}+O(\delta n^2)
\nn
&\approx \sum_i U\bar n_{i\downarrow} a^\dagger_{i\uparrow}a_{i\uparrow}+U\bar n_{i\uparrow}a^\dagger_{i\downarrow} a_{i\downarrow}-U\bar n_{i\uparrow}\bar n_{i\downarrow}
 \end{align}
 where in the last line we used the relations 
 \be
 \delta n_{i\uparrow} = a^\dagger_{i\uparrow}a_{i\uparrow} -\bar n_{i\uparrow}
 ,\quad
 \delta n_{i\downarrow} = a^\dagger_{i\downarrow}a_{i\downarrow} -\bar n_{i\downarrow}
 . 
 \ee
Replacing the Hubbard interaction in Eq.~\eqref{eq:H_main} with the above expression yields a quadratic Hamiltonian that can be diagonalized and used to construct mean-field ordered states. 

Namely, introducing local density-energies for the up and down spins, which take values $u$ and $u'$ on the two sublattices, as given in Eq.~\eqref{eq:main_uu'}, we arrive at the Hamiltonians for the spin-up and spin-down Bloch states:
\be
 H=\sum_\sigma \lp \sum_{\la ii'\ra} t a^\dagger_{i\sigma}a_{i'\sigma}
 +\sum_i u_i a^\dagger_{i\sigma}a_{i\sigma}\rp
 \ee
Single-particle energies and eigenstates can now be described by
$2\times 2$ Bloch Hamiltonians for the bipartite lattice 
\be\label{eq:hamiltonian_final}
H_\uparrow=\lp\begin{array}{cc} u & f(k) \\ f(k)^\ast & u'\end{array}\rp
,\quad
H_\downarrow=\lp\begin{array}{cc} u' & f(k) \\ f(k)^\ast & u\end{array}\rp
,
\ee
where we introduced notation $f(k)=\sum_j te^{i\vec k\vec a_j}$ (here $j$ labels vectors pointing from a given site to nearest-neighbor sites). We will use unprimed and primed variables to label the quantities associated with each of the two sublattices. Diagonalizing these matrices yields two Bloch bands
 \be
 \epsilon_\lambda(k)=\frac{u+u'}2+\lambda \sqrt{\frac{\Delta^2}{4} +|f(k)|^2}
,\quad
\lambda =\pm 1 \label{eq:energy_eigenvalues}
 ,
 \ee
 with accompanying two component eigenvectors $\lvert\psi_{s,\vec{k}}\rangle$. Here the plus/minus sign labels the conduction and valence bands. 
 
  \begin{figure}[!tb]
\centering

\includegraphics[width=\linewidth]{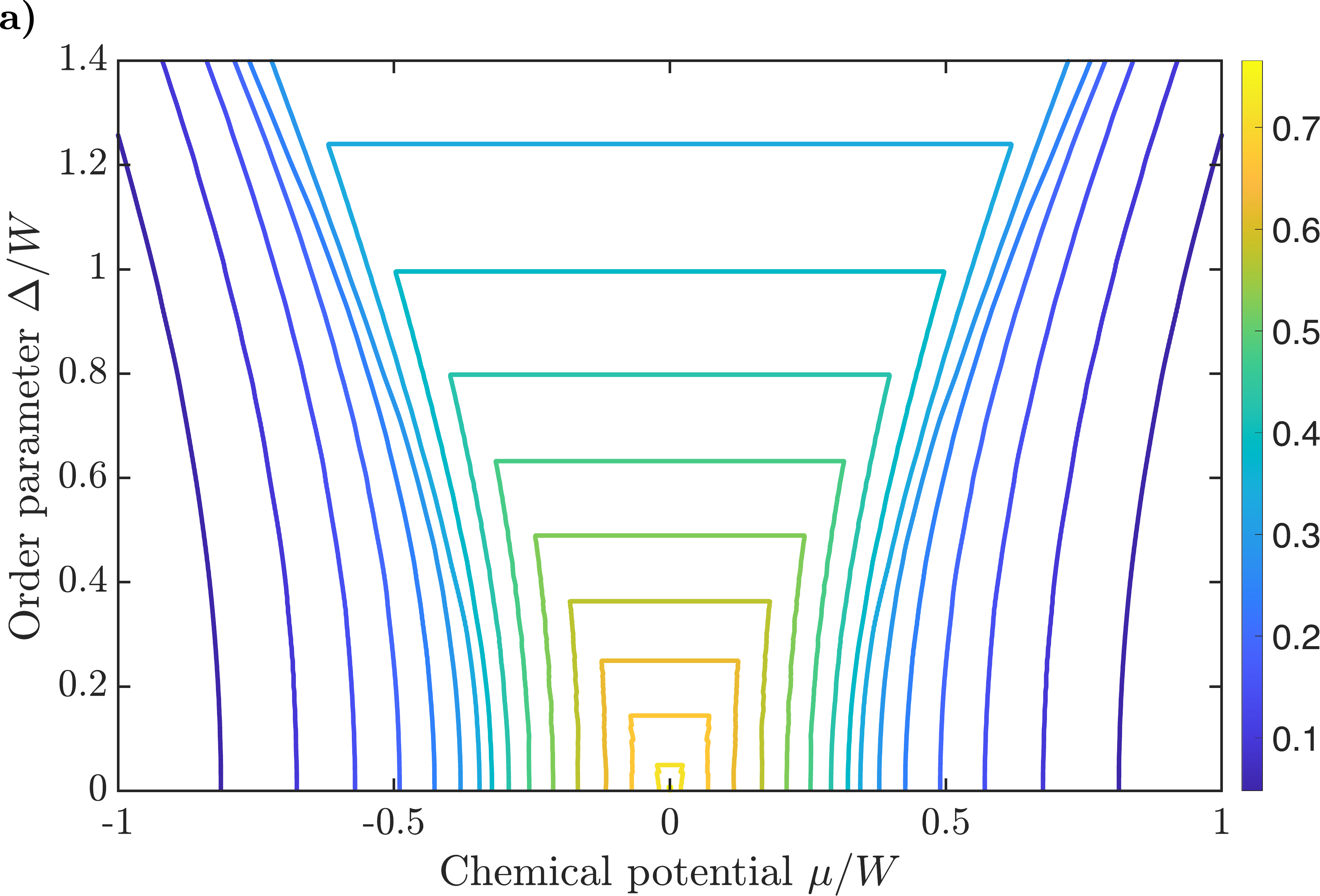}
\includegraphics[width=\linewidth]{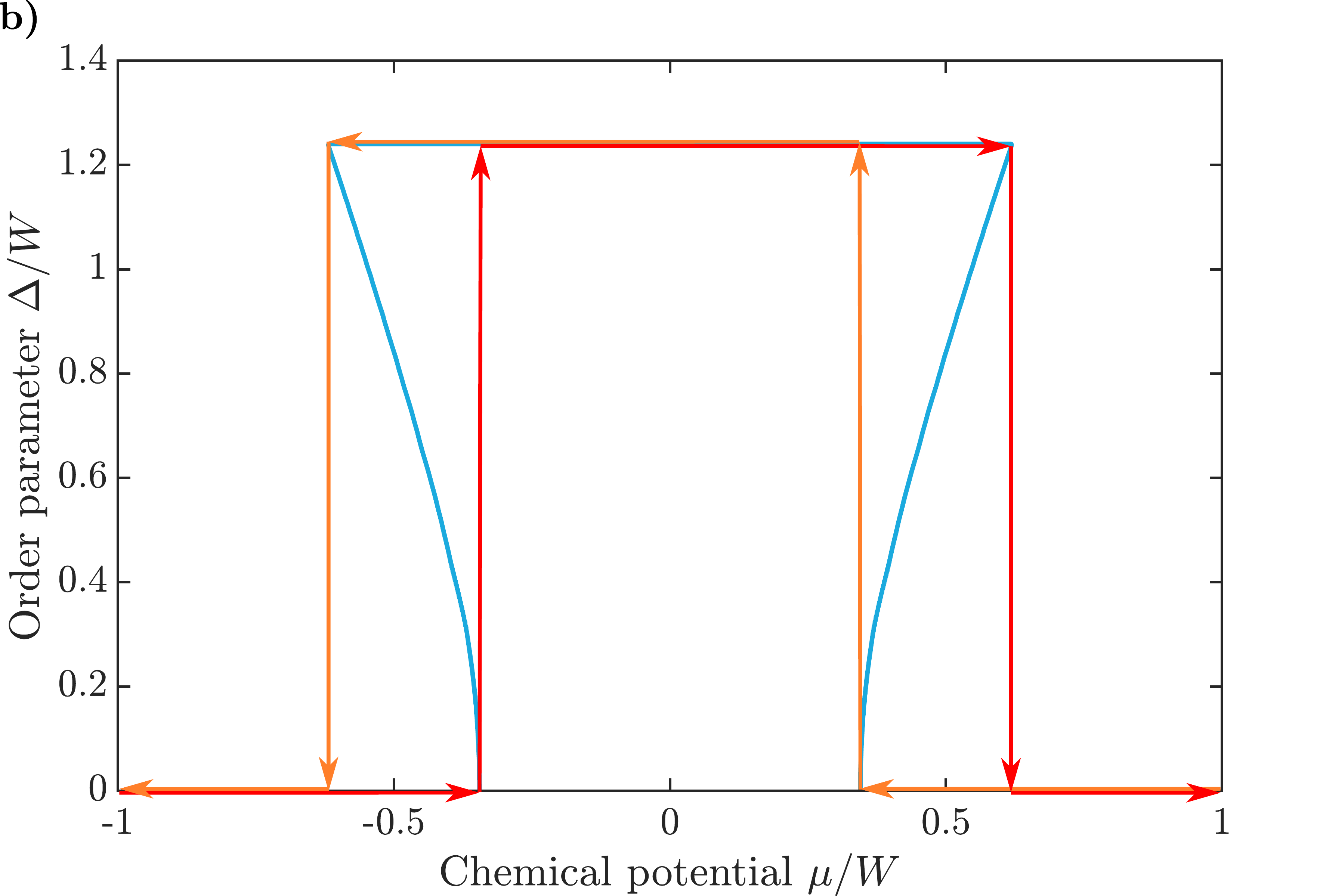}
\caption{a) Contour plots of the right-hand side of Eq.~\eqref{eq:sc_equation}. Matching its value to to $1/U$ yields the dependence of the order parameter $\Delta$ on the chemical potential $\mu$.
b) The obtained contour from panel (a) for $U = 3$ meV (blue) and the value of the order parameter (red and orange) upon sweeping the chemical potential. Arrows indicate the sweep directions.
}
\label{fig:fig_2}
\end{figure}

 The parameter $\Delta$ controls the gap between the two bands, and although it appears as a parameter in the toy-model Hamiltonian Eq.~\eqref{eq:hamiltonian_final}, its value has to be determined self-consistently at every filling density $n$. The equation which defines $\Delta$, Eq.~\eqref{eq:main_uu'}, using the eigenstates and eigenvalues of the hamiltonian $H$, takes the form
 \be
 \label{eq:sc_equation}
\frac{1}{U}=\sum_k \frac{ n_{+}(k)-n_{-}(k)}{\sqrt{\Delta^2/4 +|f(k)|^2}}\,,
\ee
where $n_\pm(k)$ corresponds to an Fermi-Dirac distribution function evaluated with either the conduction ($+$) or valence ($-$) band dispersion from Eq.~\eqref{eq:energy_eigenvalues}. In what follows we solve this integral equation self-consistently, see Fig.\ref{fig:fig_2}, to yield a dependence of $\Delta$ as a function of filling $n$ (or chemical potential $\mu$) which is then used for the analysis of collective modes. As expected from an order parameter, $\Delta$ vanishes everywhere away from the critical filling but at the critical filling it becomes non-zero.

Next, we discuss the bandstructure details of the model used in the above analysis.  We use a hexagonal-lattice tight binding model which possesses the same symmetry and the same number of subbands as the flat bands in TBG. We match the energy and length scales by choosing the width of a single band $W$ and the hexagonal lattice period $L_M$ identical to the parameters in TBG:  $W=3.75$ meV and $L_M=a/2\sin(\theta/2)$ being the moir\'e superlattice periodicity. For the magic angle value $\theta=1.05^\circ$, using carbon spacing $a=0.246$ nm, this gives $L_M=13.4$ nm. The hopping element $t$ and nearest neighbours $\vec{a}_j$ in the function $f(k)$ in the above Hamiltonian are therefore equal to $t = W/3$ and  $\vec a_j=(\cos(2\pi j/3),\sin(2\pi j/3))L_M/\sqrt{3}$, 
$j=0,1,2$.

\section{Collective modes in a Hubbard insulator}

We proceed now to find plasmon excitations of this electronic system. After introducing the necessary formalism, which explains the appearance of the new optical plasmon mode, we will divide the discussion into two sections: one dedicated to properties of the acoustic plasmon dispersion and one focusing on the new optical plasmon mode.

The acoustic and optical plasmon modes correspond to oscillations of charges in-phase and out-of-phase within the two sites of the unit cell in response to an applied electric field. As  the pattern of charge motion can be different on these two sites, the dynamical response of the electronic system that specifies their dispersion has to necessarily be different on these two sites as well. To account for this, we define a generalized polarization function explicitly projected on the A/B sublattice sites 
\be 
	\Pi(\omega, \vec q) = \lp \begin{array}{cc}
		\Pi_{AA}(\omega, \vec q) & \Pi_{AB}(\omega, \vec q) \\
		\Pi_{BA}(\omega, \vec q) & \Pi_{BB}(\omega, \vec q)
	\end{array} \rp.
\ee
Here, $\Pi_{XY}(\omega, \vec q)$ is the polarization function projected onto the sublattice sites X and Y, i.e. it is the linear response function that relates the induced charge on sublattice X to the total electric field in sublattice Y. Accordingly the dielectric function in this effective ``atomic'' site basis is 
\be \label{eq:dielectric_matrix}
	\epsilon(\omega, \vec q) = \mathbb{1} - V(\vec q) \Pi(\omega, \vec q)\,,
\ee
where $\Pi(\omega, \vec q)$ is the generalized polarization function of above and $V(\vec{q})$ accounts of Coulomb interaction between different sites (defined later). Plasmons, both the acoustic and optical branch, correspond to collective charge motion in the absence of any external perturbation, e.g. $\epsilon(\omega, \vec q) \vec{E} = 0$ where $\vec{E}$ is the total electric field in the sample. As such these two plasmon branches will arise from a solution of an equation $||\epsilon(\omega, \vec q)|| = 0$, where $||\dots||$ denotes a determinant.

We approximate the projected polarization function $\Pi_{XY}(\omega, \vec q)$ with the RPA expression\cite{mahan2000many-particle}
\begin{align} \label{eq:RPA_projected}
	\Pi_{XY}&(\omega, \vec q) = 2 \sum_{\vec k, s, s'} \frac{f_{s, \vec k + \vec q} - f_{s', \vec k}}{\epsilon_{s, \vec k + \vec q} - \epsilon_{s', \vec k} - \omega - i0^+}\times\nonumber \\
	\times & \left< X \left| \psi_{s, \vec k + \vec q} \right> \left< \psi_{s, \vec k + \vec q} \right| Y \left> \right< Y \left| \psi_{s', \vec k} \right> \left< \psi_{s', \vec k}\right| X \right>\,,
\end{align}
where the indices $X,Y$ correspond to either $A,B$ sublattices. The matrix overlaps project $X$ and $Y$ components of Bloch wavefunctions of the $\vec{k}$ and $\vec{k}+\vec{q}$ eigenstates of the Hamiltonian in Eq.\eqref{eq:hamiltonian_final}. In the above, summation  $\sum_{\vec{k}}$ denotes integration over the Brillouin zone (BZ), the composite indices $s=(\lambda,\sigma),s'=(\lambda',\sigma')$ run over the electron bands $\lambda = \pm$ and spins $\sigma = \uparrow, \downarrow$. The additional factor of $2$ in front of the summation accounts for the valley degeneracy of the toy model. Here, $f_{s,\vec k}$ is the Fermi-Dirac equilibrium distribution.

The Coulomb interaction matrix $V(\vec q)$'s elements, $V_{XY}(\vec q)$, are also Fourier transform of the Coulomb interaction between sublattice sites X and Y, i.e.
\be \label{eq:V_XY}
	V_{XY}(\vec q) = \sum_{\vec R} V(\vec R + \vec r_X - \vec r_Y) e^{-i \vec q \cdot \vec R},
\ee
where summation $\sum_{\vec R}$ runs over the Bravais lattice points in the real space, $\vec r_X$ is the position of the sublattice site X in the unit cell ($\vec r_A = \vec 0$ and $\vec r_B = \vec a_0 = (L_M/\sqrt(3), 0)$), and $V(\vec r) = 2\pi e^2 / \kappa r$ is the Coulomb interaction in a medium with a background dielectric constant $\kappa$.

While the secular equation $||\epsilon(\omega, \vec q)|| = 0$ specifies the dispersion of the two collective modes simultaneously at any wavelength, as plotted in Fig.\ref{fig:fig_1}, it is helpful to consider characteristic equation for both plasmon modes separately. To that end we define two vectors 
\be \label{eq:pm_q_def}
\lvert \pm, \vec{q}\rangle = \frac{1}{\sqrt{2}}\lp \begin{array}{c}
		1  \\
		\pm e^{i \vec{q} \cdot \vec{r}_{B}}
	\end{array} \rp.
\ee
which project the polarization components responsible for the acoustic and optical branches. Physically, these correspond to electric-field components in-phase and out-of-phase electric field established on both lattices. An electric field characterized by wavenumber $\vec q$ in BZ may have any wavenumber $\vec q + \vec G$ in the extended zone, where $\vec G$ is a reciprocal lattice vector. Such a wave will have relative phases $e^{i (\vec q + \vec G) \cdot \vec r_X}$ in the unit cell sublattices. The response to this electric field, can has an extended wavenumber $\vec q + \vec G'$, where $\vec G'$ is also a reciprocal lattice vector that as a result of pseudo-momentum conservation in the lattice, can be different than $\vec G$. 

Having defined the response function matrices, we can investigate system's response to total electric fields with different wavelengths. 

\section{Acoustic branch}

To proceed further analytically, we first rewrite \eqref{eq:RPA_projected} by performing a standard replacement $\vec{k}+\vec{q}\to-\vec{k}$ in the term containing $f_{s,\vec{k}+\vec{q}}$ followed by $-\vec{k}-\vec{q}, -\vec{k} \to \vec{k}+\vec{q}, \vec{k}$ justified by the $\vec{k}\to -\vec{k}$ symmetry of the model. This gives
\begin{nalign} \label{eq:RPA_projected_simp}
	\Pi_{XY}&(\omega, \vec q) = 4 \sum_{\vec k, s, s'} \frac{f_{s', \vec k} (\epsilon_{s', \vec k} - \epsilon_{s, \vec k + \vec q})}{(\epsilon_{s, \vec k + \vec q} - \epsilon_{s', \vec k})^2 - (\omega + i0^+)^2}\\
	\times & \left< X \left| \psi_{s, \vec k + \vec q} \right> \left< \psi_{s, \vec k + \vec q} \right| Y \left> \right< Y \left| \psi_{s', \vec k} \right> \left< \psi_{s', \vec k}\right| X \right>.
\end{nalign}

Now the long-wavelength response to a long-wavelength electric fields is obtained by putting $\vec G = \vec G' = \vec 0$ and considering a small momentum $\vec q$. In that case, the total long-wavelength charge response of the unit cell is
\be \label{eq:tot_charge_response}
	\tilde \Pi (\omega, \vec q) = \left< +, \vec q \right| \Pi (\omega, \vec q) \left| +, \vec q \right>,
\ee
where $\left| +, \vec q \right>$ was defined in Eq.\eqref{eq:pm_q_def}. Substituting the matrix elements of $\Pi_{XY}$ from \eqref{eq:RPA_projected_simp}, we get
\begin{nalign} \label{eq:dielectric_function_tb_step_2_main}
\tilde \Pi(\omega,\vec q) \simeq 4 \sum_{\vec{k},s,s'} f_{s',\vec k}\frac{F^{ss'}_{\vec{k},\vec{k}+\vec{q}}(\epsilon_{s',\vec{k}}-\epsilon_{s,\vec{k}+\vec{q}})}{(\epsilon_{s,\vec{k}+\vec{q}}-\epsilon_{s',\vec{k}})^2-(\omega+i0)^2}.
\end{nalign}
Here, $F_{\vec k + \vec q, \vec k}^{s s'} = \left| \left< \psi_{s, \vec k + \vec q} | \psi_{s', \vec k} \right> \right|^2$ describes the conventional band overlap factors. The behavior of this expression at small $\vec q$, which will be of interest for us, can be further simplified analytically.

To understand the effect correlated state has on the acoustic plasmon dispersion, it is necessary to evaluate the above long-wavelength polarization function from Eq.~\eqref{eq:tot_charge_response} in both the metallic regime where $\Delta = 0$, and the correlated regime where $\Delta \neq 0$ and the chemical potential is inside the correlation-driven gap. The sum over the band indices $s,s'$ in the polarization function from Eq.~\eqref{eq:dielectric_function_tb_step_2_main} consists of two types of terms: those corresponding to intraband transitions $s=s'$, labeled as $\Pi^{1}$, and those corresponding to interband transitions $s=-s'$, labeled as $\Pi^{2}$. When the system is in a correlated state and the chemical potential is inside the gap, then the only contribution to the polarization function comes from the interband transitions. In a metallic regime both interband and intraband transitions contribute to the polarization function with the interband contributions becoming more relevant the closer chemical potential is to the edge of the other band.

We focus first on the insulating regime as it best demonstrates the physics driving the behavior of over-the-band plasmons. When the system is correlated, the chemical potential $\mu$ lies inside the gap, c.f. the inset of Fig.\ref{fig:fig_1} and Fig.\ref{fig:fig_2}b. This translates to a condition
\be
\label{eq:temp_label_eq_1}
f_{(-,\sigma),\vec{k}} = 1\,,\quad f_{(+,\sigma),\vec{k}} = 0
\ee
on the Fermi-Dirac distribution functions which  allows us to further simplify the polarization function from Eq.~\eqref{eq:dielectric_function_tb_step_2_main}. Note that here and in what follows, for clarity, we explictly expanded the composite index $s=(\lambda,\sigma)$ labeling the band $\lambda=\pm$ and spin indices $\sigma$. As a result of Eq.\eqref{eq:temp_label_eq_1}, the polarization function contains only interband terms which gives an expression
\be
\tilde \Pi(\omega,\vec q)=4 \sum_{\vec{k},\sigma} \frac{F^{(-,\sigma)(+,\sigma)}_{\vec{k},\vec{k}+\vec{q}}(\epsilon_{(-,\sigma),\vec{k}}-\epsilon_{(+,\sigma),\vec{k}+\vec{q}})}{(\epsilon_{(+,\sigma),\vec{k}+\vec{q}}-\epsilon_{(-,\sigma),\vec{k}})^2-(\omega+i0)^2}
\,,
\label{eq:dielectric_function_tb_step_3_main}
\ee
where we used the orthogonality of the Bloch wavefunctions for different spins. The $\omega$ frequency regime relevant for the over-the-band behavior is that of a plasmon mode extending above the typical energy scale of electronic excitations. 
For simplicity of the analysis we can assume however that $\Delta \ll W$ and verify later on numerically that the qualitative behavior is not altered outside of this regime. 

The characteristic energy of interband transitions is therefore on the scale of the bandwidth $W$. In this limit the polarization function can therefore be expanded by treating $\omega$ as the largest energy scale to yield
\be
\tilde \Pi(\omega,\vec q)\approx-\frac{4}{\omega^2} \sum_{\vec{k},\sigma} F^{(-,\sigma)(+,\sigma)}_{\vec{k},\vec{k}+\vec{q}}(\epsilon_{(-,\sigma),\vec{k}}-\epsilon_{(+,\sigma),\vec{k}+\vec{q}})
.
\label{eq:dielectric_function_tb_step_4}
\ee
The above approximation to the polarization function is valid for all momenta $\vec{q}$ provided that the frequency $\omega$ is larger than that of any electronic transitions that contributed to the sum in Eq.\eqref{eq:dielectric_function_tb_step_3_main}.

In the narrow-band materials, the Fermi momentum $k_F$ is on the order of the reciprocal lattice vector and the over-the-band behavior is present for a range of momenta, see Fig.\ref{fig:fig_1}. We can therefore expand all analytic expressions in the long-wavelength limit of $q \ll k_F$. To leading order in momentum $q$ the band overlap factors therefore become
\be
\label{eq:coherence_factor_tb}
F^{(-,\sigma)(+,\sigma)}_{\vec{k},\vec{k}+\vec{q}} \approx \frac{1}{4} \sin^2\theta \frac{q^2}{k^2}+\frac{1}{16} \frac{\Delta^2 }{ v_F^2 k^2} \cos2\theta \frac{q^2}{k^2}\,,
\ee
where we approximated the non-interacting tight binding $|f(k)|^2 = v_F^2 k^2$ as a Dirac cone and introduced an angle $\theta$ between $\vec{k}$ and $\vec{q}$. We see that in the limit of $\Delta  \to 0$ it reduces to the massless graphene result. The energy difference to leading order in $q$ becomes
\be
\epsilon_{(-1,\sigma),\vec{k}}-\epsilon_{(1,\sigma),\vec{k}+\vec{q}}\approx -2 \sqrt{\frac{\Delta^2}{4}+ v_F^2 k^2}\,,
\ee
which when combined with the expression for the band overlap factors from above gives a final form for the polarization function
\begin{align}
\Pi^2(\omega,\vec q)&\approx 
\frac{2}{\pi} W \frac{q^2}{\omega^2}-\frac{1}{2\pi} \frac{\Delta^2}{W} \frac{q^2}{\omega^2}\label{eq:interband_contribution}\,.
\end{align}
The above polarisation function translates to a plasmon dispersion
\be
\omega_p  \approx \sqrt{4 \alpha W\left(1- \frac{\Delta^2}{4 W^2}\right)v_F q} \, \quad \mathrm{for}\quad \omega_p \gg W\label{eq:dispersion_mott}
\ee
which shows that the contribution of the CI gap $\Delta$ is to soften the plasmon dispersion. This behavior is to be expected as a gap between bands, here the CI gap $\Delta$, reduces the Bloch wavefunction overlap. This is in agreement with earlier studies\cite{PhysRevLett.113.246407,PhysRevB.90.235105} of plasmon modes in correlated systems of Mott type, where stronger interaction strength $U$ was shown to reduce plasmon stiffness. Most crucially however, as we will see later, presence of a correlated state also makes the dispersion independent of the chemical potential. The latter observation is again to be expected as the chemical potential is inside the CI gap and hence plasmons simply originate from transitions from a fully filled to a fully empty band. These interband plasmons are precisely the source of the over-the-band behavior seen in Fig.\ref{fig:fig_1}.

\begin{figure}[!tb]
\centering

\includegraphics[width=\linewidth]{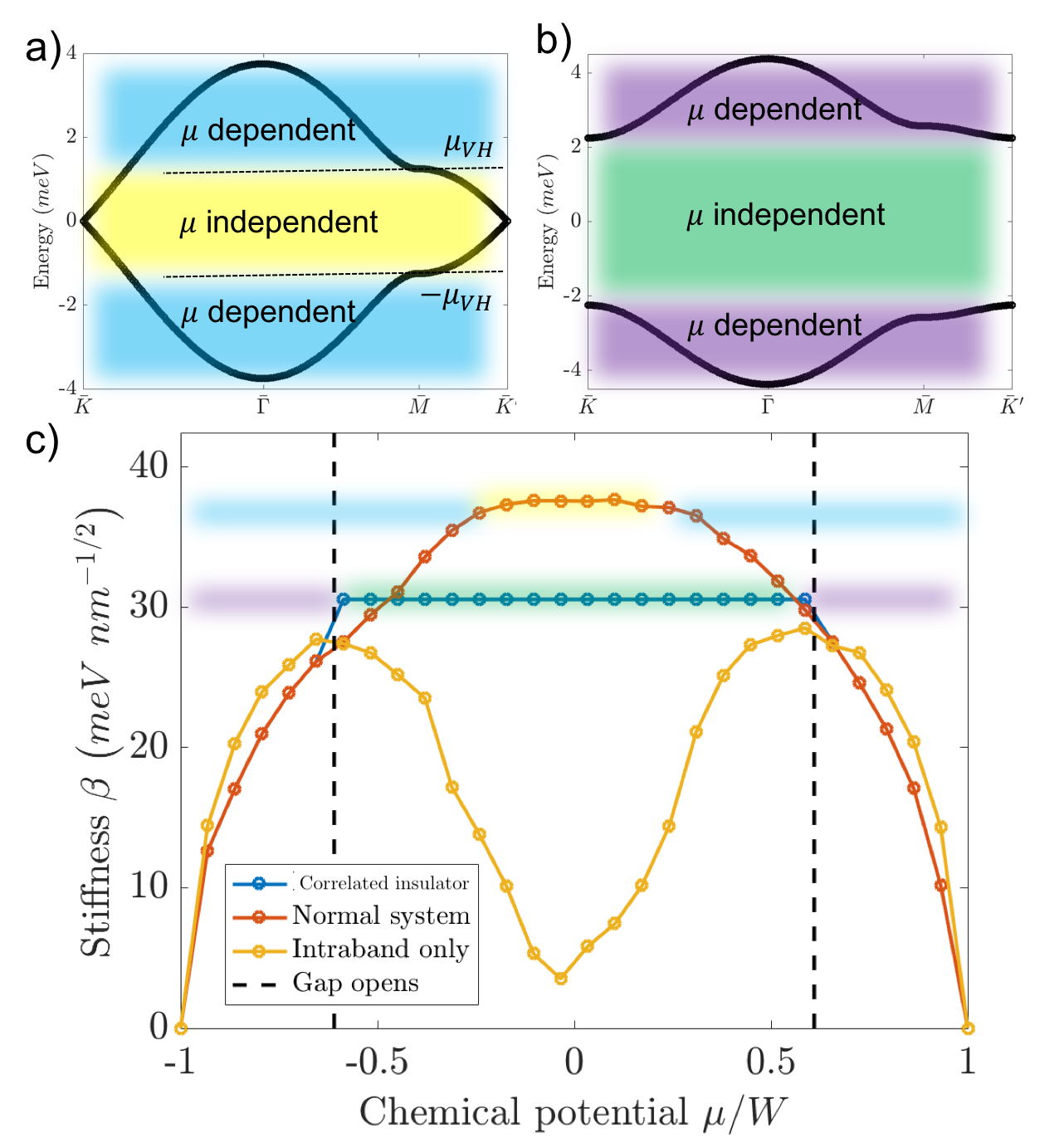}
\caption{a) Bandstructure of a normal system and its van-Hove singularity ($\mu_{VH}$). The color shaded regions correspondong to chemical dependent or independent stiffness $\beta$, Eq.~\eqref{eq:plasmon_full_normal}. b) A bandstructure in a correlated state. A CI opens a gap in the bandstructure leading to a $\mu$ independent stiffness $\beta$. c) Dependence of stiffness $\beta$ on chemical potential in a normal system, with (red line) and without (yellow line) interband transitions, and in a correlated regime (blue line). Interband transitions are behind the saturating over-the-roof behavior of plasmons in TBG. Both normal and correlated system's plasmon stiffness has regions of $\mu$-independent plasmon stiffness (see main text).}
\label{fig:fig_3}
\end{figure}

We now focus on the other limit of $\Delta = 0$ which translates to a normal system in the absence of the correlated state. The polarization function for such a system consists now of two terms that account for interband and intraband transitions. The polarization function due to intraband transitions has the simple form of\cite{mahan2000many-particle}
\be
\Pi^1(\omega,\vec q)\approx \frac{n}{m} \frac{q^2}{\omega^2} \approx \frac{2}{\pi} E_F \frac{q^2}{\omega^2}\label{eq:intraband_contribution}\,,
\ee
where $n$ is the charge density that gives rise to intraband plasmons and $m = k_F/v_F$ is the electron mass. The charge density $n$ denotes the concentration of the minority carriers in a band, either electrons or holes, as a fully empty or a fully filled band does not give rise to longitudinal charge oscillations. We denote the chemical potential corresponding to the van Hove critical point (for both positive and negative filling) as $\mu_{VH}$ (see Fig.~\ref{fig:fig_3}a). With this notation, for parameter $E_F$ we get
\be
E_F = \left\{
\begin{array}{ll}
      |\mu| & \text{ for} \,|\mu| < |\mu_{VH}|\,, \\
      W-|\mu| & \text{ for} \, |\mu| >|\mu_{VH}|\,, \\
\end{array} 
\right. , 
\ee
which amounts to measuring a chemical potential from either the bottom or the top of the band depending on whether van Hove singularity resides above or below of a corresponding filling.

As argued above, the microscopic origins behind the over-the-band modes stem from interband transitions. What controls the scale of these transitions is the Bloch wavefunction overlap Eq.~\eqref{eq:coherence_factor_tb}, which is suppressed far from the bandstructure edges. As the chemical potential comes closer to the edge of the bandstructure near $\mu = 0$, the impact of the interband transitions on the polarization function becomes maximal. An expression that captures this behavior can be derived as
\be
\Pi^2(\omega,\vec q)\approx \frac{2}{\pi} (W-|\mu|) \frac{q^2}{\omega^2}\,.\label{eq:interband_contribution_normal}
\ee
Note how the  Eq.~\eqref{eq:interband_contribution}  can be obtained from Eq.~\eqref{eq:interband_contribution_normal} by setting $\Delta = 0$ and placing the chemical potential $\mu = 0$ at the band crossing point.

Stiffness of the plasmon dispersion carriers information about the underlying electron transitions that give rise to the collective modes. Far from the other bandstructure edge, shaded blue region in the Fig. \ref{fig:fig_3}a, for $|\mu| > |\mu_{VH}|$, we expect the total polarization function to be dependent on the chemical potential $\mu$. Near the bandstructure crossing,  shaded yellow region in the Fig.~\ref{fig:fig_3}a, for $|\mu| < |\mu_{VH}|$ we expect interband and intraband contributions to the polarization function to combine and, due to the form of Eq.~\eqref{eq:intraband_contribution} and Eq.~\eqref{eq:interband_contribution_normal}, to lead to a cancellation of the dependence on the chemical potential. Plasmon dispersion $\omega_p$ therefore should follow, in a system without a Mott gap,
\be
\label{eq:plasmon_full_normal}
\omega_p(\vec{q}) = \left\{
\begin{array}{ll}
      \sqrt{8 \alpha (W-|\mu|) v_F q} & \text{ for} \,|\mu| > |\mu_{VH}|\,, \\
      \sqrt{4 \alpha W v_F q} & \text{ for} \, |\mu| < |\mu_{VH}|\,. \\
\end{array} 
\right.
\ee 
This independence of the plasmon stiffness $\beta$, defined as $\omega_p  = \beta \sqrt{q}$, on the chemical potential is a hallmark behavior of interband plasmons that are behind the over-the-band modes. This contribution of interband transitions to the plasmon stiffness $\beta$ in a normal system is shown in Fig.~\ref{fig:fig_3}c where we plot a stiffness due to both interband and intraband transitions (red line) and due to only intraband transitions (yellow line). As expected from Eq.~\eqref{eq:intraband_contribution} plasmon stiffness vanishes near the bandstructure edges.

The acoustic part of the plasmon dispersion in a correlated Hubbard insulator similarly exhibits no dependence on chemical potential in the plasmon stiffness. This is shown as a blue line in Fig.~\ref{fig:fig_3}c. As argued in Eq.~\eqref{eq:dispersion_mott} in a correlated state plasmon dispersion has no dependence on chemical potential just as the mode in the metallic state. Here however this lack of dependence on chemical potential arises simply because plasmons originate only from interband transitions as the chemical potential lies inside the gap, the green shaded region in Fig.~\ref{fig:fig_3}b. The effect of the ordered state is to soften plasmon dispersion as visible on the Fig.~\ref{fig:fig_3}c and given by Eq.~\eqref{eq:dispersion_mott}.

The chemical potential-independent plasmon stiffness of the acoustic branch in both normal and correlated Hubbard insulator has two different underlying microscopic mechanisms. One occurs due purely to inter- and intraband transition kinematics while the other is simply a property of a gap opening. It is however precisely due to the difference in these two mechanisms, and the fact that under experimental conditions the tunable parameter is charge density rather than the chemical potential, that the experimental presentation of this effect would manifest in two qualitatively different ways. In a metallic system varying chemical potential or charge density is interchangeble. We thus expect the same behavior, as that seen in Fig.~\ref{fig:fig_3}b (red curve), to also occur as a function of charge density. In a correlated insulator the relation between chemical potential and filling is more complicated. A correlation-driven insulating behavior which spans a whole sample occurs only precisely at the integer filling, in this model corresponding to $\mu = 0$. Yet, according to the calculation shown in Fig.~\ref{fig:fig_2}b, a gap can open far from the $\mu = 0$ value. This is because as chemical potential is brought closer to $\mu = 0$, then correlated domains of increasing spatial size are formed giving rise to a gap opening in that correlated region. Collective modes are therefore sensitive to the local Hamiltonian, i.e. whether it has a gap or not, provided that the wavelength of the plasmon is smaller than the correlation length. In a near-field optical microscopy measurement\cite{basov2012,koppens2012} a moving tip at various positions of a sample and global chemical potential could therefore be used as a probe of whether a sample region possesses a gap for an expected metallic filling - a hallmark behavior of a correlated insulator. 

\section{Optical branch}\label{sec:sec_optical}

\begin{figure}[!tb]
	\includegraphics[width=\linewidth]{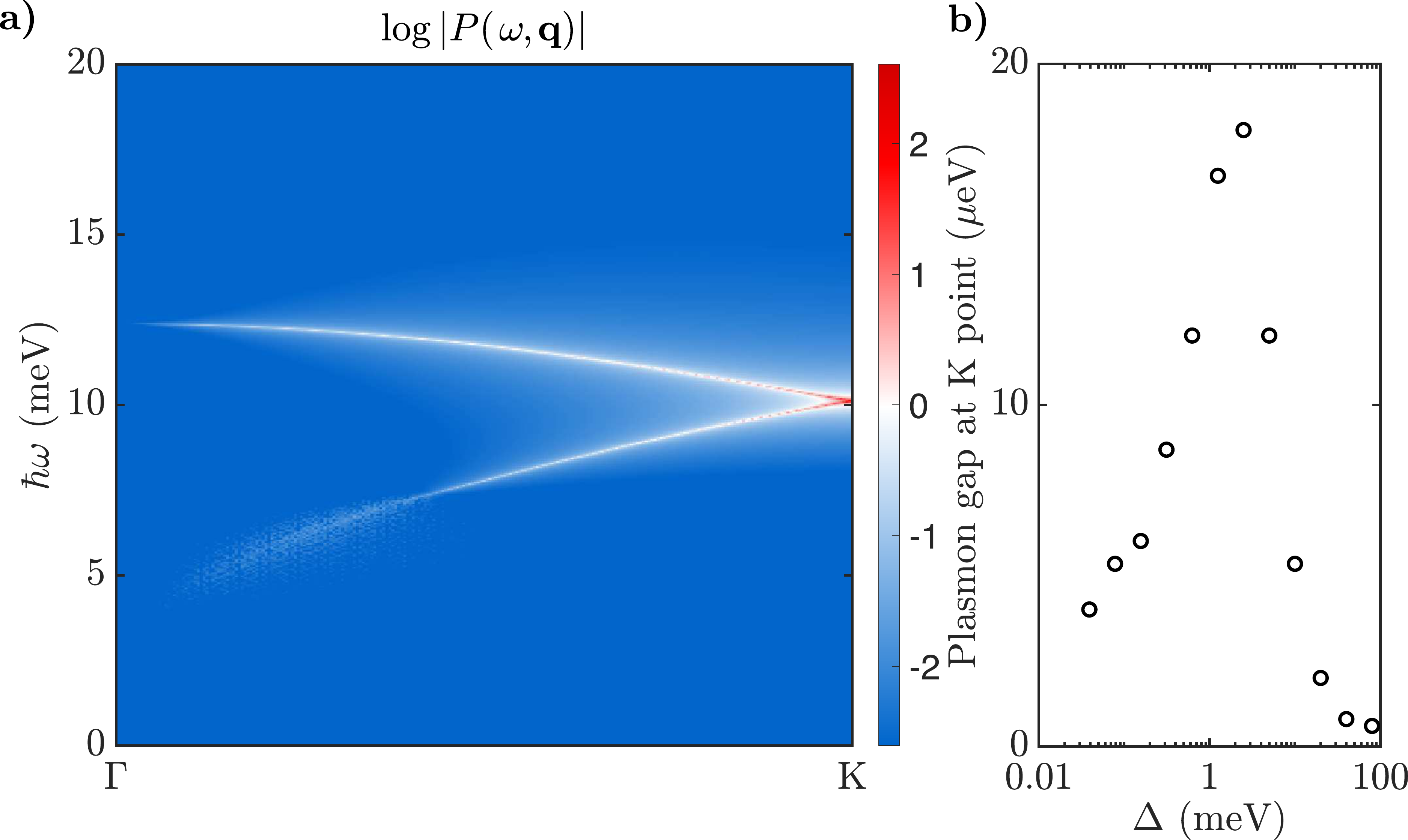}
	\caption{a) The dipole response of plasmon bands of a narrow-band toy model with a correlated Hubbard insulator. The optical branch shows especially large dipole moment as a result of correlation-driven sublattice asymmetry. Parameter values are chosen to mimic TBG bands (bandwidth $W = 3.75$~meV, lattice periodicity $L_M=13.4$~nm, Fermi energy is placed inside the gap at $\mu = 0$).
	b) The small gap between the acoustic and optical plasmon branches at corners of BZ as a function of induced gap in electron bandstructure $\Delta$.}\label{fig:fig_4}
\end{figure}

We focus now on the new optical branch of the plasmon dispersion shown in Fig.~\ref{fig:fig_1}. The existence of this branch is guaranteed as having two sites per unit cell in the real space lattice results in two plasmon branches in the reciprocal space. One significance of having an optical mode is the presence of the finite frequency plasmons as plasmon wavenumber approaches $\Gamma$-point. However, the experimental relevance of these plasmons depends on the dipole moment of the modes.

In order to analyze this, we define two optical responses. First, the charge response
\begin{equation}\label{eq:charge_response}
	Q(\omega, \vec q) = 	\frac{\left< +, \vec q \left | \Pi(\omega, \vec q) \right| +, \vec q \right>}{||\epsilon(\omega, \vec q)||},
\end{equation}
and then, the dipole response
\begin{equation}\label{eq:dipole_response}
	P(\omega, \vec q) = 	\frac{\left< -, \vec q \left | \Pi(\omega, \vec q) \right| +, \vec q \right>}{||\epsilon(\omega, \vec q)||}.
\end{equation}
Vectors $\left| \pm , \vec q \right>$ are defined in Eq.~\eqref{eq:pm_q_def}.

Both quantities are constructed from $\Pi(\omega, \vec q)\left| +, \vec q \right>$, which is the charge distribution on two sublattices in response to an applied electric field with a unit magnitude and wavenumber $\vec q$. If we simply sum the charges of the two sublattices, after canceling the anticipated relative phase corresponding to the wavenumber $\vec q$, we obtain the total induced charge. This gives the numerator of the charge response, $Q(\omega, \vec q)$, in Eq.~\eqref{eq:charge_response}. If instead of the total induced charge, we like to get the difference of the charge induced on two sublattices, or the ``dipole moment'', we can subtract the components of $\Pi(\omega, \vec q)\left| +, \vec q \right>$ after canceling the relative phase due to the finite wavenumber. We can do this by finding the inner product of it with $\left< -, \vec q \right|$ that gives the numerator of the dipole response, $P(\omega, \vec q)$, in Eq.~\eqref{eq:dipole_response}. Finally, in order to see the response for the plasmon modes, we divide both quantities by $||\epsilon(\omega, \vec q)||$ that vanishes along the modes.

The charge response shows the total optical activity of the acoustic plasmon branch. But when an optical branch exists as a result of presence or emergence of a sublattice structure, we expect the dipole response to be nonzero for this new mode. If the sublattice structure emerges only after an A/B gap is induced by correlations, the dipole response of the optical branch will go to zero when we take the induced gap, $\Delta$, to zero. But if the sublattice structure already exists in the A/B symmetric lattice, like the hexagonal lattice in our toy model, we expect the dipole moment to be nonzero for the optical branch even when $\Delta=0$. However, we see that emergence of an A/B asymmetry will further enhance the dipole response of the optical branch. In Fig.~\ref{fig:fig_4}a, we show the dipole response for a hexagonal lattice with a correlation-driven A/B sublattice asymmetry. The parameters are similar to the plasmon modes in Fig.~\ref{fig:fig_1}.

Another feature of the optical branch, is the gap between this branch and the acoustic one at the corners of BZ. When $\Delta=0$ and the hexagonal lattice has A/B symmetry, plasmon modes form a gapless plasmonic Dirac cone around K point in BZ\cite{stauber2020Dirac}. But with a finite $\Delta$, this is no longer required and this plasmonic Dirac cone can be gapped. Indeed, as we previously saw in Fig~\ref{fig:fig_1}, a small gap is opened for this case. In Fig.~\ref{fig:fig_4}b, we show the magnitude of this gap as a function of $\Delta$.

\section{Conclusion}

Collective oscillations of narrow-band materials exhibit a striking over-the-band behavior which decouples them from the spectrum of underlying single-particle excitation energies. The origins of this separation of energy scales can be traced back to a strong Coulomb coupling of electron-hole pairs in these systems. 

While the over-the-band behavior is a universal feature present in all narrow-band materials, the salient details of the dispersion of the collective modes are dependent on the nature of the correlated state. Correlated behavior leading to a correlation-driven band gap opening softens the dispersion of collective modes. This softening of the dispersion as a function of a filling number (or a local chemical potential) can be used in near-field optical microscopy measurements as a local probe of a correlated state.

Most crucially however the appearance of a correlated behavior gives rise to a emergence of an effective sublattice. This sublattice results in halving of a moir{\'e}-BZ and folding of the plasmon dispersion. Resulting optical plasmon branch has a finite frequency at zero momentum. These new collective optical modes poses a strong dipole moment making them more easily detectable in optical microscopy measurements.

\bibliography{references}

\end{document}